\documentclass[prl,aps,twocolumn,superscriptaddress,showpacs]{revtex4}
\usepackage[dvips]{graphicx,color}\usepackage[dvips]{graphicx,color}
\def\(({\left(}
\def\)){\right)}
\def\[[{\left[}
\def\]]{\right]}
\def\nle{\ \raise.3ex\hbox{$<$}\kern-0.8em\lower.7ex\hbox{$\sim$}\ }
\def\nge{\ \raise.3ex\hbox{$>$}\kern-0.8em\lower.7ex\hbox{$\sim$}\ }

\begin{document}
\title{
Scaling Analysis of Domain-Wall Free-Energy in the Edwards-Anderson 
Ising Spin Glass in a Magnetic Field}

\author{M. Sasaki}
\affiliation{
Department of Applied Physics, Tohoku University, Sendai, 980-8579, Japan}

\author{K. Hukushima}
\affiliation{
Department of Basic Science, University of Tokyo, Tokyo, 153-8902, Japan}

\author{H. Yoshino}
\affiliation{
Department of Earth and Space, Osaka University, Toyonaka, 560-0043, Japan}

\author{H. Takayama}
\affiliation{
Institute for Solid State Physics, University of Tokyo, Kashiwa, 277-8581, Japan}

\date{\today}

\begin{abstract}
The stability of the spin-glass phase against a magnetic field is studied 
in the three and four dimensional Edwards-Anderson Ising spin glasses. 
Effective couplings $J_{\rm eff}$ and effective fields $H_{\rm eff}$ 
associated with length scale $L$ are measured by a numerical domain-wall 
renormalization group method. The results obtained by scaling analysis 
of the data strongly indicate the existence of a crossover length beyond 
which the spin-glass order is destroyed by field $H$. 
The crossover length well obeys a power law of $H$ 
which diverges as $H \rightarrow 0$ but remains finite for any non-zero $H$, 
implying that the spin-glass phase is absent even in an 
infinitesimal field. These results are well consistent with the droplet 
theory for short-range spin glasses. 

\end{abstract}
\pacs{75.10.Nr, 75.40.Mg, 05.10.Ln}

\maketitle
In spite of extensive studies for more than two decades, a basic problem 
on the field-temperature phase diagram of the short-range Ising Spin Glass (SG)
is still controversial. The mean-field theory predicts the existence 
of the SG phase in a magnetic field up to a certain strength 
at a temperature below $T_{\rm c}$, the critical temperature 
in a zero field~\cite{AlmeidaThouless78}. On the other hand, 
the droplet theory~\cite{FisherHuse88,BrayMoore84,BrayMooreHC}, 
a phenomenological theory for short-range SGs, predicts 
the absence of the SG phase even in an infinitesimal field. 

In experiments, this issue has been addressed by the study of Ising SG 
${\rm Fe}_x{\rm Mn}_{1-x}{\rm TiO}_3$. 
Although the presence of the SG phase in field was 
first reported~\cite{KatoriIto94}, the subsequent work concluded 
its absence by careful analyses of its relaxation time~\cite{Mattsson95}. 
The same conclusion was recently drawn 
by J\"onsson {\it et al}~\cite{Jonsson05}. 
From a theoretical point of view, numerical studies of 
the Edwards-Anderson (EA) short-range SG 
model~\cite{EdwardsAnderson75} have yielded rather conflicting results: 
some data support the presence of the SG phase 
in field~\cite{Marinari98,Krzakala01}, and others support its absence~\cite{HoudayerMartin99,TakayamaHukushima04,YoungKatzgraber04}. 
However, a recent numerical work on the correlation length has shown 
the absence of the SG phase in three dimensions 
at a low field $H=0.05J$~\cite{YoungKatzgraber04}, 
where $J$ is the standard deviation of 
couplings $\{ J_{ij} \}$. Since this field is much below a critical field 
$H_{\rm c}\approx 0.65J$ suggested by a previous study~\cite{Krzakala01}, 
the result strongly indicates the absence of the SG phase. 
However, there still remains the possibility that the SG phase 
is stable at lower fields. Furthermore the physical 
mechanism which  destroys the SG state remains to be clarified.

In the present work, we study the EA SG in field 
in both three and four dimensions by the same
numerical Domain-Wall Renormalization-Group (DWRG) method that we have
recently used in studying the fragility of the SG equilibrium state
against small changes in temperature 
or in bonds $\{ J_{ij} \}$~\cite{Sasaki05}. 
%\textcolor{red}{It should be noted that both the dimensions are 
%below the upper critical dimension of the EA SGs, $d_{\rm u}=6$.} 
We also show some corresponding results of the Migdal-Kadanoff (MK) SG 
model, in which the absence of the SG phase in field has already
been shown~\cite{MiglioriniBerker98}. 
For this model we can easily access huge sizes such as 
$L\approx 10^{10}$. This compensates the Monte Carlo (MC) results on the EA 
model within a limited range of length scales. 
Quite interestingly, both the models exhibit the same scaling behavior. 
All the observables are scaled as a function of $L/\ell_{\rm cr}(H)$, where 
$L$ is the system size and $\ell_{\rm cr}(H)$ a crossover length beyond 
which the SG order is destroyed by field $H$. The existence of 
such a length scale is indeed predicted by the droplet theory. 
The observed scaling behavior indicates the absence of the SG phase 
even in an infinitesimal field.

{\it The droplet theory---}
Let us begin with a brief survey of the droplet 
theory~\cite{FisherHuse88,BrayMoore84,BrayMooreHC}.
Droplets are spin clusters which can be flipped with a low excitation energy. 
The typical excitation energy of droplets with length scale 
$\ell$ is assumed to scale as $\ell^{\theta}$, where $\theta$ is the so-called 
stiffness exponent. Now let us consider to apply field $H$. 
Although we consider a uniform field for simplicity, 
the following argument is also valid for random fields. 
In Ising SGs, the spins in a droplet are either $+1$ or $-1$ 
with equal probability, implying that the order of Zeeman energy of droplets 
with size $\ell$ is $H\ell^{d/2}$. 
Since the droplet theory provides some arguments which support the inequality 
$\frac{d-1}{2}\ge \theta$~\cite{FisherHuse88}, the theory claims that 
there exists a characteristic length, given as 
$\ell_{\rm cr}(H)\approx H^{-1/\zeta}$ with $\zeta\equiv d/2-\theta$, 
that droplets larger than $\ell_{\rm cr}(H)$ are forced to flip by the field. 
As a result, the SG state at $H=0$ becomes unstable
beyond $\ell_{\rm cr}(H)$. 
We call $\ell_{\rm cr}$ and $\zeta$ the (field) 
crossover length and the crossover exponent, respectively. Furthermore 
the droplet theory claims that the system is paramagnetic beyond 
$\ell_{\rm cr}(H)$ as it happens in random field Ising model. 
Because $\ell_{\rm cr}(H)$ diverges as $H\rightarrow 0$ 
but remains finite for any non-zero $H$, the droplet theory asserts 
the absence of the SG phase in any field.

\begin{figure}[b]
%\begin{center}
   \includegraphics[width=7.6cm]{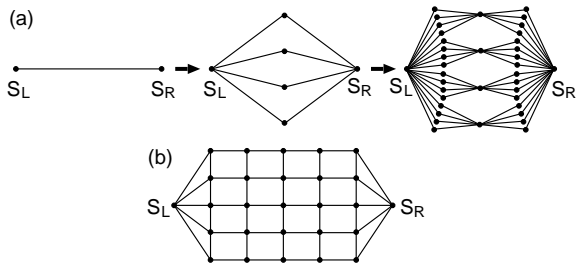}
%\end{center}
  \caption{(a) The construction of the hierarchical lattice ($b=2$, $d=3$).
(b) The lattice for the DWRG method in the EA SG. 
}
\label{fig:MKlattice}
\end{figure} 

{\it DWRG method and observables---}
Let us first explain models for the DWRG 
method~\cite{BrayMoore85,Hukushima99,Sasaki05}. 
Figure~\ref{fig:MKlattice}~(a) shows the way to construct the hierarchical 
lattice for the MK SG. The lattice is made iteratively by replacing 
each bond with $b^{d-1}$ paths, where $d$ is the dimension of the lattice. 
Each path consists of $b$ bonds, and new $(b-1)$ spins (full circles) are 
inserted in between. We hereafter call the two outermost spins ($S_{\rm L}$ 
and $S_{\rm R}$) {\it boundary spins}. 
The size $L$ of the lattice is multiplied by $b$ as the replacement is done 
once. Figure~\ref{fig:MKlattice}~(b) is the lattice for the EA SG. 
The lattice is the same as that in ref.~\cite{Sasaki05}. 
It consists of two boundary spins and $L^d$ spins 
on a $d$-dimensional hyper-cubic lattice. 
The boundary condition is open in the direction along which 
the boundary spins lie, and is periodic in other directions. 
The Hamiltonian is 
\begin{equation}
{\cal H}=-\sum\nolimits_{\langle ij \rangle}J_{ij} S_i S_j-\sum\nolimits_{i} H_i S_i,
\end{equation}
where the first term is exchange energies between two nearest neighboring spins and the second term is Zeeman energies 
by field $H_i$. 

In the DWRG method, we measure the effective coupling $J_{\rm eff}$ 
and the effective fields $H_{\rm eff}$ defined by
\begin{eqnarray}
Z(S_{\rm L},S_{\rm R})&\equiv& {\rm Tr'} {\rm e}^{-H\{S\}/T}\propto 
{\rm e}^{{-\mathcal H}_{\rm eff}(S_{\rm L},S_{\rm R})/T},
\label{eqn:E_Hamiltonian1}\\
{\mathcal H}_{\rm eff}(S_{\rm L},S_{\rm R})&=&-J_{\rm eff}S_{\rm L} S_{\rm R} 
-H_{\rm eff}^{\rm (L)} S_{\rm L}-H_{\rm eff}^{\rm (R)} S_{\rm R},
\label{eqn:E_Hamiltonian2}
\end{eqnarray}
where the trace in Eq.~(\ref{eqn:E_Hamiltonian1}) is for all the spins 
except $S_{\rm L}$ and $S_{\rm R}$. In the MK SG, 
$Z(S_{\rm L},S_{\rm R})$ is estimated exactly by taking the trace 
sequentially from the later generated spins to the earlier generated ones. 
For a detailed description of the procedure, we refer the reader 
to~\cite{MiglioriniBerker98}. 
In the EA SG, on the other hand, probability 
$P(S_{\rm L},S_{\rm R})$ is measured by MC 
simulations~\cite{Sasaki05}.
Since $J_{\rm eff}$ and the free-energy difference 
$\delta F$ caused by twisting the two boundary spins are related 
by $J_{\rm eff}=-\delta F/2$ in zero field, we consider that 
$J_{\rm eff}$ represents the strength of the SG order.
Because $J_{\rm eff}$ is either positive or negative, we calculate 
the standard deviation of sample-to-sample fluctuations of 
effective couplings, $\sigma_{\rm J}(L,H)$. 
We also measure that of effective fields $\sigma_{\rm H}(L,H)$. 
The correlation between effective couplings in zero field and those 
in field $H$ is also estimated by the correlation coefficient
\begin{equation}
C(L,H)=\frac{\overline{J_{\rm eff}(L,H)J_{\rm eff}(L,0)}}
{\sigma_{\rm J}(L,H)\sigma_{\rm J}(L,0)},
\label{eqn:CF}
\end{equation}
where $\overline{\cdots}$ is the sample average. Here $J_{\rm eff}(L,H)$ and $J_{\rm eff}(L,0)$ are calculated for the same realization of $\{ J_{ij} \}$.

\begin{figure}[b]
   \includegraphics[width=8.4cm]{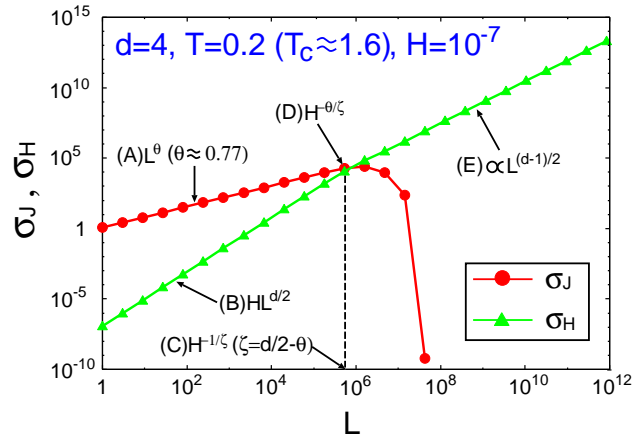}
  \caption{Size dependences of $\sigma_{\rm J}$ and $\sigma_{\rm H}$
in the four dimensional MK SG 
at $T\approx 0.125T_{\rm c}$, 
where $T_{\rm c}$ is the critical temperature in zero field. 
Strength of random fields is $10^{-7}$. 
(A) Size dependence of $\sigma_{\rm J}$ for smaller $L$. 
(B) Size dependence of $\sigma_{\rm H}$ for smaller $L$. 
(C) The size where the crossover between $\sigma_{\rm J}$ and 
$\sigma_{\rm H}$ occurs.
(D) The value of $\sigma_{\rm J}$ and $\sigma_{\rm H}$ at the crossing point. 
(E) Size dependence of $\sigma_{\rm H}$ for larger $L$. 
}
\label{fig:MKcrossing}
\end{figure}

\begin{figure}[b]
%\begin{center}
  \includegraphics[width=8.4cm]{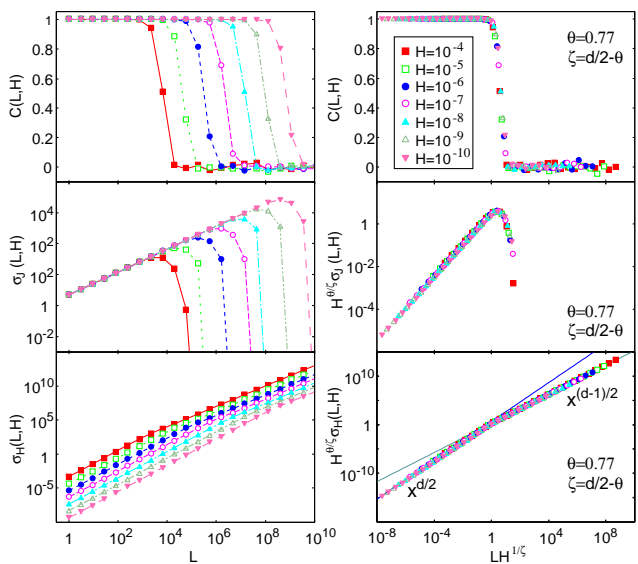}
%\end{center}
  \caption{Raw data (left panel) and their scaling plots (right panel) 
in the MK SG. The dimensions and the temperature are the same as 
those in Fig.~\ref{fig:MKcrossing}. 
%The fields examined are 
%shown in the legend of the right-bottom figure. 
The value $\theta=0.77$ obtained in ref.~\cite{NifleHilhorst93} is used 
for the scaling. The two straight lines in the right-bottom panel are 
proportional to $x^{d/2}$ and $x^{(d-1)/2}$, respectively.}
\label{fig:MKresult}
\end{figure}

{\it Results in the MK SG---}
Figure~\ref{fig:MKcrossing} shows the size dependences of $\sigma_{\rm J}$ and $\sigma_{\rm H}$ in the four-dimensional MK SG. 
The values of couplings $\{J_{ij}\}$ are  
taken from a Gaussian distribution of mean $0$ and width $1$. 
We apply random fields $H_i$ of strength $H$ by following the way 
in~ref.~\cite{MiglioriniBerker98}. 
The pool method~\cite{BanavarBray87} is used to access huge sizes 
such as $L\approx 10^{10}$. 
The Zeeman energy $(\sigma_{\rm H}\propto HL^{d/2})$ 
overwhelms the effective coupling $(\sigma_{\rm J}\propto L^{\theta})$ 
around $L\approx H^{-1/\zeta}$, i.e., around the crossover length 
$\ell_{\rm cr}(H)$ in the droplet theory. 
After the crossing, $\sigma_{\rm J}$ exhibits roughly exponential decay 
and the exponent of $\sigma_{\rm H}$ changes from $d/2$ to $(d-1)/2$. 
These observations naturally lead us to the idea that $H^{\theta/\zeta}
\sigma_{\rm J}$, $H^{\theta/\zeta} \sigma_{\rm H}$ and $C$ are scaled as
functions of $X \equiv LH^{1/\zeta}$ which we call the scaling variable. 
The idea is tested in Fig.~\ref{fig:MKresult}. 
The data with different $H$ and $L$ 
nicely collapse into scaling curves. 
These results clearly show the existence of the crossover length 
$\ell_{\rm cr}(H)$.

{\it Results in the EA SG---}
Let us first explain some details of our simulation. The values of $J_{ij}$ 
are either $+J$ or $-J$ with equal weights. Uniform field $H$ is applied 
to all the spins except $S_{\rm L}$ and $S_{\rm R}$. 
The number of samples is more than $1,500$ for all the sets of $(L,H)$. 
We use the exchange MC method~\cite{HukushimaNemoto96} to accelerate 
the equilibration, and the method in~ref.~\cite{Hukushima99} 
to overcome a hard-relaxing problem of the boundary spins which is 
originated from their high connectivities. The temperature ranges used 
for the exchange MC are $0.5J\le T\le 4.0J$ 
($T_{\rm c} \approx 1.1J$~\cite{Katzgraber06}) 
for $d=3$ and $1.0J\le T\le 4.5J$ 
($T_{\rm c} \approx 2.0J$~\cite{MarinariZuliani99}) 
for $d=4$. 
We hereafter focus on the data at the lowest temperature 
which is well below $T_{\rm c}$.
We set the MC step for thermalization and that for measurement to be equal. 
They are sufficiently (at least 5 times) larger than the ergodic time to ensure the equilibration, 
where the ergodic time is the average MC step for a specific replica to move
from the lowest to the highest temperature and return to the lowest one. 
As done in ref.~\cite{Hukushima99}, we have also checked that MC runs 
starting from parallel and anti-parallel boundary spins yield 
identical results within error bars.

\begin{figure}[b]
%\begin{center}
  \includegraphics[width=8.4cm]{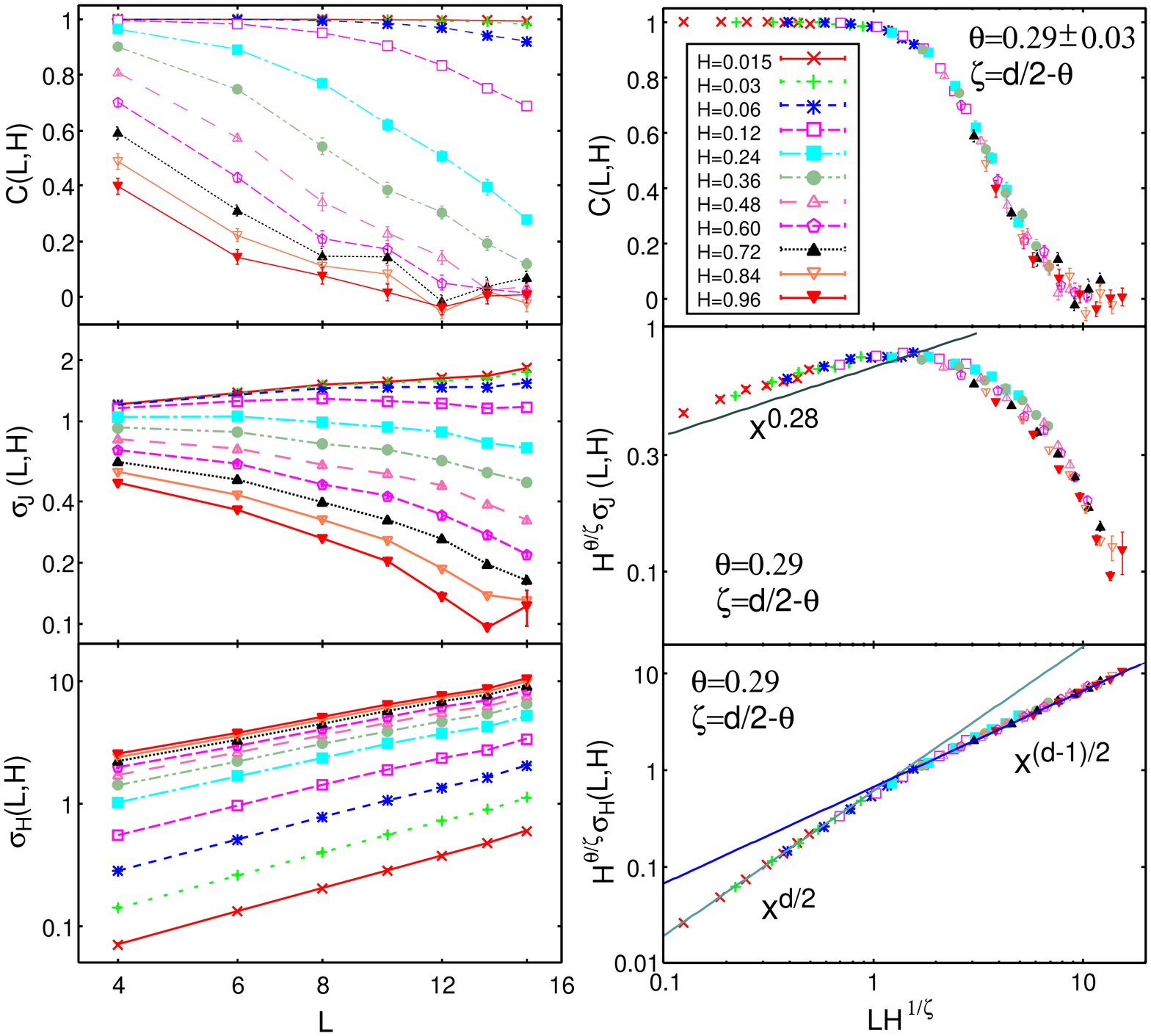}
%\end{center}
  \caption{
Raw data (left panel) and their scaling plots 
(right panel) in the EA SG at $T=0.5J$ for $d=3$. The value of $\theta$ is 
estimated by fitting the data of $C$, and the value is used for the scaling 
plot of $\sigma_{\rm J}$ and that of $\sigma_{\rm H}$. 
The slope of the line in the right-middle panel is obtained by 
linear least-square fits of $\ln[\sigma_{\rm J}(L,H=0)]$ against $\ln(L)$.}
\label{fig:3dEAresult}
\end{figure}

\begin{figure}[h]
%\begin{center}
  \includegraphics[width=8.4cm]{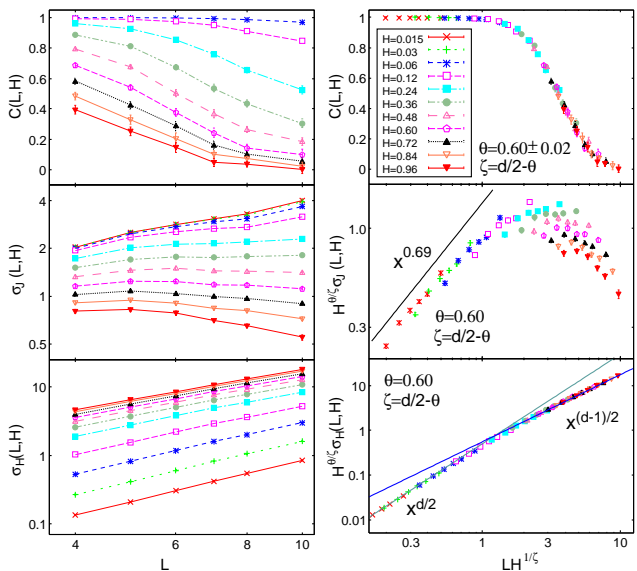}
%\end{center}
  \caption{
The same plot as that in Fig.~\ref{fig:3dEAresult} 
for the four dimensional EA SG at $T=1.0J$.}
\label{fig:4dEAresult}
\end{figure}

Figures~\ref{fig:3dEAresult} and~\ref{fig:4dEAresult} show the results 
for $d=3$  and those for $d=4$, respectively. 
Since $C$ is a dimensionless quantity, it is a function of only 
$X=LH^{1/\zeta}$.
We therefore estimate $\zeta$ by fitting the data of $C$. 
By assuming the scaling relation $\zeta=d/2-\theta$ 
predicted by the droplet theory, $\theta$ is estimated to be $0.29(3)$ 
for $d=3$  and $0.60(2)$ for $d=4$. Since they are close to both recent 
estimations by Boettcher~\cite{Boettcher04} 
(0.24(1) for $d=3$  and 0.61(2) for $d=4$) and direct estimations by linear 
least-square fits of $\ln[\sigma_{\rm J}(L,H=0)]$ against $\ln(L)$ 
(0.28(3) for $d=3$  and 0.69(3) for $d=4$), our data 
support the scaling relation.

We next examine scaling properties
of $\sigma_{\rm J}$ and $\sigma_{\rm H}$ by using the values of $\theta$
determined above. The scaling reasonably works except 
$\sigma_{\rm J}$ for $d=4$. The deviation suggests 
that the fields investigated are too large or/and the sizes are 
too small so that corrections to the scaling is not negligible. 
In fact, if we closely observe the scaling plot of 
$\sigma_{\rm J}$ for $d=3$, we notice that the data with large $H$, 
say $H\ge 0.48$, systematically deviate from the master curve, while 
the data with small field are scaled quite well. 
If we estimate $\theta$ by forcing all the data to be
scaled {\it approximately}, we get apparently larger 
values of $\theta$ in both $\sigma_{\rm J}$ and $\sigma_{\rm H}$. 
For example, $\theta$ estimated from $\sigma_{\rm J}$ 
in such way is $0.36\pm 0.02$ for $d=3$ and $0.84\pm 0.04$ for $d=4$. 

The crossover behavior in the MK SG shown in Fig.~\ref{fig:MKresult} 
look very sharp in comparison with that in the EA SG. 
This is simply because the ranges of $H$ and $L$, and so $X$, 
are quite wide. The enlargement of the crossover region such as to 
$0.1 \nle X \nle 10$ is realized by choosing the parameter ranges, 
for example, $0.015J \le H \le 0.95J$ and $L=3^n$ with $1 \le n \le 4$.
These ranges are close to those we are forced to choose 
for the EA SG. 
We then obtain quite similar results (not shown) as those obtained in the EA
SG (Fig.~\ref{fig:3dEAresult}). 
Within these ranges $\sigma_{\rm J}$ increases monotonically 
for smaller $H$ and it decreases monotonically for larger $H$.
Still, such data are reasonably scaled with the same value of $\theta$ 
as indicated in Fig.~\ref{fig:MKresult}. These observations
imply that the observed scaling behaviors in both the MK and 
EA SGs are essentially the same. 

%We first find that the scaling in $\sigma_{\rm H}$ works well 
%in both dimensions. 
%We also notice from the data of $\sigma_{\rm J}$ for $d=3$ 
%that the data with small field, say $H\le 0.36$, are scaled nicely 
%while the data with large $H$ systematically deviate from the scaling. 
%A similar tendency is observed for $d=4$ although 
%the scaling is worse than that for $d=3$. Lastly, we comment 
%a possible reason why the scaling of $\sigma_{\rm J}$ is valid 
%for small $H$, whereas that of $C$ seems to holds for all $H$. 
%By substituting $J_{\rm eff}(H)=J_{\rm eff}(0)+aH^2+bH^4+\cdots$ 
%(note that $J_{\rm eff}(H)$ is an even function of $H$) with the defining 
%equations of $C$ and $\sigma_{\rm J}$, we can easily find that 
%the leading term 

%The values of $\theta$ are estimated by scaling fit of each observables.
%The estimated values are a bit higher than the previous estimations. 
%For example, the estimations in ref.~\cite{Boettcher04} are 0.24(1) 
%for three dimensions and 0.61(2) for four dimensions. 
%\textcolor{red}{This is probably due to effect of correction to scaling 
%which is relevant for these small system sizes.} However, we see that 
%the scaling works nicely. The scaling functions 
%in both the MK and EA spin glasses are rather similar as discussed 
%further in detail below. 

{\it Interpretation of results---}
We first consider the crossover behavior in $\sigma_{\rm H}$. 
Since the field is not applied to the boundary spins, the effective fields
$H_{\rm eff}^{\rm (L)}$ and $H_{\rm eff}^{\rm (R)}$ originate only through the 
influence of the field applied to the bulk spins $\{S_i\}$. 
For example, if the correlation between $S_i$ and $S_{\rm L}$ is positive, 
an upward field to $S_i$ tends to direct $S_{\rm L}$ upwards. 
Since the correlation can be either positive or negative depending on the
site, the contribution to the effective fields can also be either positive or 
negative. When the effective coupling $J_{\rm eff}$ exists, 
$S_{\rm L}$ and $S_{\rm R}$ receive such random contributions, 
whose amplitude is proportional to $H$, from all the bulk spins. 
This yields $\sigma_{\rm H}$ which is proportional to $HL^{d/2}$. 
When $J_{\rm eff}$ vanishes, on the other hand, 
$\sigma_{\rm H}\propto L^{(d-1)/2}$ because 
the boundary spins, which interact with $L^{d-1}$ spins, 
receive contributions only from spins around their surfaces.

Now the meaning of our results are clear. 
As shown in Fig.~\ref{fig:MKcrossing}, the Zeeman energy ($\sigma_{\rm H}$)  begins to overwhelm the effective coupling ($\sigma_{\rm J}$) around $L\approx \ell_{\rm cr}(H)$ 
since $\sigma_{\rm H}$ increases faster than $\sigma_{\rm J}$. 
After that, $\sigma_{\rm J}$ rapidly drops to zero. 
This means that the SG order whose length scale is 
longer than $\ell_{\rm cr}(H)$ is destroyed by field. 
As described above, $\sigma_{\rm H}$ is kept increasing but with the change of its exponent from $d/2$ to $(d-1)/2$.
The decay of $C(L,H)$ to zero for larger scaling variable $X$ 
indicates the vanishing of the correlation between the state 
in zero field and that in field.
%Since $X$ in the thermodynamic limit is infinite for any non-zero $H$ and
%$\sigma_{\rm J}$ is zero there, our results 
The observed scaling behavior consistently implies the absence 
of the SG phase in any non-zero field.

%\begin{figure}[h]
%%\begin{center}
%  \includegraphics[width=8.4cm]{Fig5C.eps}
%%\end{center}
%  \caption{
%The same plot as that in Fig.~\ref{fig:3dEAresult} 
%for the four dimensional EA SG at $T=1.0J$.}
%\label{fig:4dEAresult}
%\end{figure} 

{\it Discussion and conclusions---}
Now let us argue on the fragility of the SG state 
against other perturbations. 
According to the droplet theory, 
a change in the temperature (or bonds) 
by the amount $\Delta$ gives rise to a new SG state 
which is decorrelated to the original one 
beyond the length scale $\ell_{\rm ch}(\Delta)$. 
Here $\ell_{\rm ch}(\Delta)$ is proportional to $\Delta^{-1/\zeta^*}$ 
with $\zeta^* \equiv d_{\rm s}/2-\theta$, $d_{\rm s}$ being 
the fractal dimension of droplets. 
This type of the fragility of the SG state is called
temperature or bond chaos~\cite{BrayMoore87}. 
Our recent DWRG study~\cite{Sasaki05} in the EA SG has indeed 
revealed the existence of $\ell_{\rm ch}(\Delta)$. 
A key observable in such studies 
is the correlation coefficient $C$ 
of Eq.~(\ref{eqn:CF}) with $H$ replaced by $\Delta$. 
As shown in Figs.~3, 4 and~5, the system exhibits 
similar fragility against the field perturbation. 
However, it must be noted that 
vanishing of $\sigma_{\rm J}$ for larger $X$ 
strongly indicates that the magnetic field destroys the
SG order itself in sharp contrast to the cases of temperature and 
bond perturbations.

%Two comments are in order. 
%One is on the behavior of the observables in the crossover region 
%of $X \sim 1$. Their crossover behavior in the MK SG shown 
%in Fig.~\ref{fig:MKresult} looks very sharp. 
%This is simply because the ranges of $H$ and $L$, and so $X$, 
%are quite wide. The enlargement of the crossover region such as to 
%$0.1 \nle X \nle 10$ is realized by choosing the parameter ranges, 
%for example, $0.015J \le H \le 0.95J$ and $L=3^n$ with $1 \le n \le 4$.
%These ranges are close to those we are forced to choose 
%for the EA SG. 
%We then obtain quite similar results (not shown) as those obtained in the EA
%SG (Fig.~\ref{fig:3dEAresult}). 
%Within these ranges $\sigma_{\rm J}$ increases monotonically 
%for smaller $H$ and it decreases monotonically for larger $H$.
%Still, such data are well scaled with the value of $\theta$ 
%in Fig.~\ref{fig:MKresult}. These observations
%imply that the observed scaling behaviors in both the MK and 
%EA SGs are essentially the same. 

%We further note that the length scale of domains 
%involved in phenomena observed by real experiments is also very 
%limited due to the extremely slow domain growth.  
%Therefore, to reach a correct answer to the basic problem of the present
%interest from the experimental side, it is necessary to interpret
%measured data by fully taking into account such 
%\textcolor{red}{slow domain growth}, 
%as demonstrated in refs.~\cite{TakayamaHukushima04,TakayamaHukushima07}.

To conclude, we have studied the EA SG in field 
by a numerical DWRG method. The thermodynamic observables of 
the EA SG are confirmed to follow the scaling 
in terms of the crossover length $\ell_{\rm cr}(H)$ as predicted 
by the droplet theory, the consequence of which is the absence of 
the SG phase in field. It should be noted that 
above the upper critical dimension $d_{\rm u}=6$ different scenario 
may hold~\cite{KatzgraberYoung05}. 
We consider that all of our results concerning the temperature, bond 
and field perturbations provide us a strong evidence for the appropriateness 
of the droplet theory 
for the description of SGs 
in low dimensions.

{\it Acknowledgments---}
This work is supported by Grant-in-Aid for Scientific 
Research Program (\#18740226 and \#18079004) from MEXT in Japan. 
The computation in this work has been done 
using the facilities of the Supercomputer Center, Institute for Solid State 
Physics, University of Tokyo.

\end{document}